\documentclass[preprintnumbers,prd,twocolumn,showpacs,floatfix,nofootinbib,superscriptaddress]{revtex4}
\usepackage{graphicx}
\usepackage{epsfig}
\usepackage{bm}
\usepackage{amssymb}
\usepackage{float}
\usepackage{amsmath}
\usepackage{dcolumn}
\usepackage{cancel}
\usepackage[colorlinks]{hyperref}
\usepackage[usenames,dvipsnames]{color}
\hypersetup{
     breaklinks=true,
    pdfstartview={FitH},    % fits the width of the page to the window
    colorlinks=true,       % false: boxed links; true: colored links
    linkcolor=blue,          % color of internal links
    citecolor=red,        % color of links to bibliography
    filecolor=magenta,      % color of file links
    urlcolor=blue,           % color of external links
    anchorcolor=green,      % Color for anchor text
    linktocpage=true
}

\newcommand{\Mpl}{M_{\textrm{Pl}}}
\renewcommand{\(}{\left(}
\renewcommand{\)}{\right)}

\newcommand{\nn}{\nonumber}
\def\al{\alpha}
\def\bet{\beta}

\def\sig{\sigma}
\def\lam{\lambda}
\def\ep{\epsilon}

\def\N{\mathcal{N}}

\def\del{\delta}

\def\doi{http://doi.org}
\def\arxiv{http://arxiv.org/abs}
 \def\t{\tilde}
 \def\e{\mathrm{e}}

\begin{document}

\title{Evading Lyth bound in models of quintessential inflation}

\author{Md. Wali Hossain}
 \email{wali@ctp-jamia.res.in}
\affiliation{Centre for Theoretical Physics, Jamia Millia Islamia,
New Delhi-110025, India}

\author{R. Myrzakulov}
\email{rmyrzakulov@gmail.com} \affiliation{ Eurasian  International
Center for Theoretical Physics, Eurasian National University, Astana
010008, Kazakhstan}

\author{M.~Sami}
\email{sami@iucaa.ernet.in} \affiliation{Centre for Theoretical
Physics, Jamia Millia Islamia, New Delhi-110025, India}

\author{Emmanuel N. Saridakis}
\email{Emmanuel\_Saridakis@baylor.edu}
 \affiliation{Physics
Division, National Technical University of Athens, 15780 Zografou
Campus,  Athens, Greece} \affiliation{Instituto de F\'{\i}sica,
Pontificia Universidad de Cat\'olica de Valpara\'{\i}so, Casilla
4950, Valpara\'{\i}so, Chile}

\begin{abstract}
Quintessential inflation refers to an attempt to unify inflation and late-time
cosmic acceleration using a single scalar field. In this letter we consider
two different classes of quintessential inflation, one of which is based upon
a Lagrangian with non-canonical kinetic term $k^2(\phi)\partial^\mu \phi
\partial_\mu \phi$ and a steep exponential potential while the second
class uses the concept of steep brane world inflation. We show that in both
cases the Lyth bound can be evaded, despite the large tensor-to-scalar ratio
of perturbations. The post-inflationary
dynamics is consistent with nucleosynthesis constraint in these cases.
\end{abstract}


\maketitle

%%%%%%%%%%%%%%%%%%%%%%%%%%%%%%%%%%%%%%%%%%%%%%%%%%%%%%%%%%%%%%%%%%%%%%%%%%%%%%%%%%%%%%%%%%%
\section{Introduction}
%%%%%%%%%%%%%%%%%%%%%%%%%%%%%%%%%%%%%%%%%%%%%%%%%%%%%%%%%%%%%%%%%%%%%%%%%%%%%%%%%%%%%%%%%%%

It is remarkable that inflationary paradigm may be directly tested
using relic gravity waves that are quantum mechanically generated
during inflation. The primordial tensor perturbations leaves its
important imprints on cosmic microwave background radiation. The
recent measurements on B-mode polarizations reveal that tensor-to-scalar
ratio of perturbations is large and inflation took place
around the GUT scale \cite{Ade:2014xna}. The large ratio of
perturbations, namely, $r\gtrsim 0.1$ imply that in case of a single
field model, the field range over which inflation took place
satisfies the Lyth bound$-$ $\delta \phi\gtrsim 5 \Mpl$
\cite{Lyth:1996im,Lyth:1998xn,Efstathiou:2005tq,Easther:2006qu,Krause:2007jr,
Baumann:2011ws,Dimastrogiovanni:2013pr,Adshead:2013qp,Hebecker:2013zda,
Aravind:2014axa,Antusch:2014cpa,German:2014qza,Gao:2014pca,Garcia-Bellido:2014eva}.
This means that inflaton potential should not be very flat allowing
a large excursion of field during inflation. It clearly rules out
small field models with $\delta \phi <\Mpl$ \cite{Antusch:2014cpa} in the standard framework.
The large field inflation in the super-Planckian region throws a big
challenge for model building in effective field theoretic context.

Models of quintessential inflation are of particular interest in
cosmology \cite{Peebles:1999fz,Sahni:2001qp,Sami:2004xk,Copeland:2000hn,Huey:2001ae,
Majumdar:2001mm,Dimopoulos:2000md,Sami:2003my,Dimopoulos:2002hm,Hinterbichler:2013we,Hossain:2014xha}.
These models aim to unify inflation with late time cosmic
acceleration using a single field. Models of quintessential
inflation broadly fall into two categories depending upon the behavior of their potentials
 during inflation: (1)  Models  with  shallow potentials  at early times and
steep thereafter and (2) those  in which inflaton potentials are
steep throughout the history of universe but turn  shallow only at late
times. In the first case,  one needs an extra mechanism to
invoke late time features in the potential.

The models of class (1): Inflation in this case can be implemented by 
field Lagrangian with non canonical kinetic term 
 and a 
steep exponential potential \cite{Hossain:2014coa,Hossain:2014xha,Wetterich:2013aca,
Wetterich:2013jsa,Wetterich:2013wza,Wetterich:2014eaa}.
In second case, inflation can take place by invoking the extra
brane damping \cite{Sahni:2001qp,Sami:2004xk,Copeland:2000hn,Huey:2001ae,
Majumdar:2001mm,Maartens:1999hf,Apostolopoulos:2005ff}. As the field rolls
down its potential and  the high energy brane corrections to
Einstein equations on the brane \cite{Shiromizu:1999wj} disappear, graceful exit from
inflation takes place. 

 In both the cases, steep behaviour after
inflation is required for the commencement of radiative regime. It
is possible to find a class of models for non-canonical case which 
might give rise to large
value of $r$ consistent with BICEP2 \cite{Hossain:2014coa} (see Refs.~\cite{Kallosh:2014xwa,Choi:2014dva,Giovannini:2014hsa,
Kappl:2014lra,Dai:2014jja,McDonald:2014oza,Cai:2014xxa,Bamba:2014mua,
Giovannini:2014bba,Xu:2014laa,Qiu:2014nla,Bousso:2014jca,
Smith:2014kka,Lyth:2014yya,Cheng:2014ota,Bonvin:2014xia,Okada:2014lxa,Choudhury:2014kma,
Gerbino:2014eqa,Gong:2014cqa,Cheng:2014bma,Miranda:2014wga,
Kobayashi:2014jga,Kehagias:2014wza,Ma:2014vua,Harigaya:2014sua,
Hazra:2014jka,Hazra:2014aea,Dinda:2014zta,Flauger:2014qra,Martin:2014lra,Barranco:2014ira,
Mortonson:2014bja,Maity:2014dsa,Garrison-Kimmel:2014kia,Mahajan:2014daa,Wan:2014fra,
Giovannini:2014jca,Ferrara:2014ima} 
on the related theme).

 In this letter
we shall examine the Lyth bound in case of models of quintessential inflation 
of both the aforesaid categories and check whether the bound can be evaded in 
compliance with the nucleosynthesis constraint.

%%%%%%%%%%%%%%%%%%%%%%%%%%%%%%%%%%%%%%%%%%%%%%%%%%%%%%%%%%%%%%%%%%%%%%%%%%%%%%%%%%%%%
\section{Lyth bound}
%%%%%%%%%%%%%%%%%%%%%%%%%%%%%%%%%%%%%%%%%%%%%%%%%%%%%%%%%%%%%%%%%%%%%%%%%%%%%%%%%%%%%%
In the following discussion we shall first review the Lyth bound in the
standard case and then we will examine it for non-canonical fields  and
models of steep braneworld inflation.
The bound indeed gets modified in these cases and the influence of post
inflationary dynamics on it is crucial.
%%%%%%%%%%%%%%%%%%%%%%%%%%%%%%%%%%%%%%%%%%%%%%%%%%%%%%%%%%%%%%%%%%%%%%%%%%%%%%
\subsection{Canonical scalar field}
%%%%%%%%%%%%%%%%%%%%%%%%%%%%%%%%%%%%%%%%%%%%%%%%%%%%%%%%%%%%%%%%%%%%%%%%%%%%%%

In case of single canonical scalar field model, the slow-roll parameters 
are defined as
\begin{eqnarray}
\label{eq:eps_can}
\epsilon&=&\frac{\Mpl^2}{2}\(\frac{1}{V}\frac{{\rm d}V}{{\rm d}\phi}\)^2 \, ,\\
\eta &=& \frac{\Mpl^2}{V}\frac{{\rm d^2}V}{{\rm d}\phi^2}\ ,
\label{eq:eta_can}
\end{eqnarray}
where $\phi$ is the canonical scalar field. In this case, the
number of e-folds is given by
\begin{eqnarray}
\label{Lbound}
\mathcal{N} &=& \frac{1}{\Mpl^2}\int_{\phi_{\rm end}}^{\phi_{\rm in}}\frac{V(\phi)}{V'(\phi)}{\rm d}\phi
\equiv \frac{1}{\Mpl}\int_{\phi_{\rm end}}^{\phi{\rm in}}
{\frac{{\rm d}\phi}{\sqrt{2\epsilon}}} \, ,
\end{eqnarray}
which along with the
field range of inflation is related to the slow-roll parameter in the
following way:
\begin{eqnarray}
\label{Lbound}
\mathcal{N}  \lesssim\frac{|\phi_{\rm in}-\phi_{\rm end}|}{\Mpl\sqrt{2
\epsilon_{\rm min}}} \, .
\end{eqnarray}

Assuming then the monotonous behaviour of the slow-roll parameters, we have $\epsilon_{\rm min}\approx\epsilon_{\rm in}$, where $\ep_{\rm in}$ is the value of $\ep$ where inflation commences. Hence, using the consistency relation $r_\star=16\epsilon_{\rm in}$, where $r_\star$ is the tensor to scalar ration at the commencement of inflation, provides us the Lyth bound,
\begin{equation}
\delta \phi\equiv {|\phi_{\rm in}-\phi_{\rm
end}|}\gtrsim\mathcal{N}\Mpl\left(\frac{r_\star}{8}\right)^{1/2},
\label{eq:lyth_bound}
\end{equation}
which implies that in the case $r_\star\gtrsim 0.1$ and $\N=50$, the field range of
inflation is given by $\delta \phi\gtrsim 5 \Mpl$
\cite{Antusch:2014cpa}. This super-Planckian field range of inflation
is of great theoretical concern.

The above quoted result is valid for single-field inflation models with a
canonical kinetic term. However, the Lyth bound is modified in case of a
single-field inflation with non-canonical kinetic term  or canonical kinetic
term with brane corrections. In the following subsections we shall examine
the Lyth bound in these two backgrounds. 

%%%%%%%%%%%%%%%%%%%%%%%%%%%%%%%%%%%%%%%%%%%%%%%%%%%%%%%%%%%%%%%%%%%%%%%%%%
\subsection{Non-canonical scalar field}
%%%%%%%%%%%%%%%%%%%%%%%%%%%%%%%%%%%%%%%%%%%%%%%%%%%%%%%%%%%%%%%%%%%%%%%%%%

Let us consider a non-canonical scalar field $\phi$, which can be transformed
to canonical form under the  transformation
\begin{equation}
 \(\frac{{\rm d}\sig}{{\rm d}\phi}\)^2=k^2(\phi) \, ,
 \label{eq:canonicalisation}
\end{equation}
where $k^2(\phi)$ is the coefficient of the kinetic term of the non-canonical 
scalar field and $\sig$ represents the canonical scalar field. Now using the transformation (\ref{eq:canonicalisation}) we 
have
\begin{equation}
 \frac{{\rm d}V(\sig)}{{\rm d}\sig}=\frac{1}{k(\phi)}\frac{{\rm d}V(\phi)}{{\rm d}\phi} \, .
\end{equation}

The slow-roll parameters for the non-canonical scalar field are then defined
as
\begin{eqnarray}
\label{eq:eps_non_can}
\epsilon&=&\frac{\Mpl^2}{2}\(\frac{1}{V}\frac{{\rm d}V}{{\rm d}\sig}\)^2
=\frac{\Mpl^2}{2k^2(\phi)}\(\frac{1}{V}\frac{{\rm d}V}{{\rm d}\phi}\)^2\, ,\\
\eta &=& \frac{\Mpl^2}{V(\sig)}\frac{{\rm d^2}V(\sig)}{{\rm d}\sig^2}
=2\epsilon(\phi)-\frac{\Mpl}{\alpha}\frac{{\rm d}\ep(\phi)}{{\rm d}\phi}\ .
\label{eq:eta_non_can}
\end{eqnarray}
In this case, the number of e-folds ($\N$) is given by
\begin{eqnarray}
 \N &=& \frac{1}{\Mpl^2}\int_{\sig_{\rm end}}^{\sig_{\rm in}}\frac{V(\sig)}{{\rm d}V(\sig)/{\rm d}\sig}{\rm d}\sig \, ,
 \nn \\ &=&\frac{1}{\Mpl^2}\int_{\phi_{\rm end}}^{\phi_{\rm in}}k^2(\phi)\frac{V(\phi)}{V'(\phi)}{\rm d}\phi \, ,
\end{eqnarray}
which   using Eq.~(\ref{eq:eps_non_can}) gives us the bound
\begin{eqnarray}
 \N \lesssim & \frac{\del\phi}{\Mpl^2} \Big|k^2(\phi) \frac{V(\phi)}{V'(\phi)}\Big|_{\rm max}
 =\frac{\del\phi}{\Mpl}\frac{k_{\rm max}}{\sqrt{2\ep_{\rm min}}}\, .
 \label{eq:lb_non_can}
\end{eqnarray}

Similarly to the previous subsection we can consider the case where $\ep_{\rm
min}\approx\ep_{\rm in}$ and the 
tensor-to-scalar ratio is $r_\star=16\ep_{\rm in}$, which after using 
relation (\ref{eq:lb_non_can}) gives us the following expression
\begin{eqnarray}
 \del\phi\gtrsim \Bigg(\N\Mpl\sqrt{\frac{r_\star}{8}}\Bigg)\frac{1}{k_{\rm max}} \, .
 \label{eq:lb_non_can1}
\end{eqnarray}
  From Eq.~(\ref{eq:lb_non_can1}) we deduce that the Lyth bound 
depends upon the value of the coefficient of the kinetic term at 
the beginning of inflation. Note that for large values of $k_{\rm max}$, we 
can have sub-Planckian Lyth bound even if we have large value 
of the tensor-to-scalar ratio $r_\star$. 

Let us now consider a particular case with the following action in the standard FRW
cosmology\cite{Hossain:2014xha,Wetterich:2013aca,Wetterich:2013jsa,
Wetterich:2013wza,Wetterich:2014eaa}:\footnote{Non-canonical field is also considered in Refs.~\cite{Nojiri:2005pu,Capozziello:2005tf,Nojiri:2006be} for unification of phantom inflation and late-time cosmic acceleration.}
\begin{eqnarray}
\label{action}
 &&\mathcal{S} = \int d^4x \sqrt{-g}\bigg[-\frac{\Mpl^2}{2}R+
\frac{k^2(\phi)}{2}\partial^\mu\phi\partial_\mu \phi+V(\phi) \bigg],\nn \\
\label{eq:action1}\\
&&k^2(\phi) = \(\frac{\al^2-\t\al^2}{\t\al^2}\)\frac{1}{1+\bet^2
\e^{\al\phi/\Mpl}}+1 \, ,
\label{eq:k}
\end{eqnarray}
where $V(\phi)=\Mpl^4\e^{-\al\phi/\Mpl}$; the parameter
 $\tilde{\alpha}$ controls the slow roll such that
$\tilde{\alpha}\ll 1$  and $\beta$ is related to the scale of
inflation. In the region where $\phi$ is large, the kinetic function
$k(\phi)\to 1$ reduces the action to the scaling  form. Nucleosynthesis
constraints \cite{Ade:2013zuv} implies that $\alpha$ should be large, thereby
the potential is steep in the post inflationary regime
\cite{Hossain:2014xha,Hossain:2014coa,Wetterich:2013jsa}.
Let us also mention that in the small-field approximation, the potential in
terms of the canonical field $\sigma\simeq \alpha \phi/\tilde{\alpha}$ has
the form of a shallow exponential potential, namely $V(\sigma)\sim
\e^{-\tilde{\alpha}\sigma/\Mpl}$ as $\t\al\ll 1$
\cite{Hossain:2014xha}. Thus, a model is suitable for quintessential
inflation provided that we invoke late-time features in the potential.

In this case, the slow-roll parameter in terms of the  non-canonical
field $\phi$ is given by \cite{Hossain:2014coa,Wetterich:2013jsa}
\begin{equation}
\epsilon=\frac{\Mpl^2}{2
k^2(\phi)}\left(\frac{1}{V}\frac{dV}{d\phi}\right)^2=\frac{\alpha^2}{2
k^2(\phi)}\to \frac{\alpha^2}{2}\left(1+X\right)
\end{equation}
where $X\equiv \beta^2\e^{\al\phi/\Mpl}$. The number
of e-folds and the slow-roll parameter is related as \cite{Hossain:2014coa}
\begin{equation}
\label{epN}
 \ep=\frac{\t\al^2}{2}\frac{1}{1-\e^{-\t\al^2\N}} \, .
\end{equation}
The large field region $X\gg 1$ corresponds to $\tilde{\alpha} ^2\ll
1/\mathcal{N}=1$ implying that that $X_{\rm end}\simeq
2/\tilde{\alpha}^2\gg 1$ which means that inflation ends in the
region of large values of the field. The boundary of small and large
field limits is given by $\tilde{\alpha}^2=1/\mathcal{N}$
\cite{Wetterich:2013jsa,Hossain:2014coa}. Note that Eq.~(\ref{epN}) implies that
$\epsilon$ is a monotonously increasing function of
$\tilde{\alpha}$. Therefore, if inflation begins in the region around the
boundary, we might improve upon the values of $\tilde{\alpha}$ and
hence the ratio $r_\star$, allowing for larger range of inflation.
Finally, using  Eq.~(\ref{eq:lb_non_can1}) we obtain the following bound for the
action~(\ref{eq:action1}):
\begin{eqnarray}
 \del\phi\gtrsim \Bigg(\N\Mpl\sqrt{\frac{r_\star}{8}}\Bigg)\frac{\t\al}{\al} \, .
 \label{eq:lb_non_can2}
\end{eqnarray}

Let us now explicitly check whether we can really get the required range
consistent with Lyth bound.  We consider
the following ratio \cite{Hossain:2014coa}
\begin{equation}
\label{Vratio}
 \frac{V_{\rm end}}{V_{\rm in}}=\frac{\tilde{\alpha}^2}{2\(
\e^{\tilde{\alpha}^2\mathcal{N}}-1\)}=\frac{r_\star}{16}\e^{-\t\al^2\N} \, ,
\end{equation}
which gives
\begin{eqnarray}
 \frac{\al}{\Mpl}|\phi_{\rm in}-\phi_{\rm end}|=
 \frac{\al\del\phi}{\Mpl}&=&\Bigg|\ln \left[\frac{\t\al^2}
{2\(\e^{\t\al^2\N}-1\)}\right]\Bigg|\nn \\
&=&\Big|\ln\(\frac{r_\star}{16}\)-\t\al^2\N\Big| \, .
\label{eq:lb_vg}
\end{eqnarray}
Then using Eq.~(\ref{epN}) and the relation $r_\star=16\ep_{\rm in}$, we find 
that $r_\star\approx0.15$ for $\t\al=0.06$ and $\N=60$. 
Considering these values and using Eq.~(\ref{eq:lb_vg}), we arrive at the 
estimate $\del\phi/\Mpl\approx5/\al$. The parameter $\al$ can be fixed from
the post inflationary dynamics using nucleosynthesis constraints, obtaining
$\al\geq 20$ \cite{Hossain:2014xha}. For $\al=20$, $\del\phi=0.25\Mpl$ which is the maximum value of $\del\phi$. However,  using 
Eq.~(\ref{eq:lb_non_can2}) and taking $\N=60$, $r_\star=0.15$, $\al=20$ and 
$\t\al=0.06$, we obtain the bound $\del\phi\geq0.0246\Mpl$. Thus, we
conclude that the model under consideration obeys Lyth bound. 

  Now, concerning the parameter $\bet$ in (\ref{eq:k}), that is related with
the scale of inflation, we can fix it from COBE normalization, which gives
us the relation 
between the parameters $\t\al$, $\bet$ and e-foldings $\N$
\cite{Hossain:2014coa} as
\begin{eqnarray}
 \frac{\bet^2\sinh^2\(\t\al^2\N/2\)}{\t\al^2}=6.36\times 10^{-8} \, .
 \label{eq:al_bet_N}
\end{eqnarray}
We can use $\bet$ from   (\ref{eq:al_bet_N}) in order to acquire the scale
of 
inflation as \cite{Hossain:2014coa}
\begin{equation}
 V_{\rm in}^{1/4}=\(\frac{2.5\times 10^{-7}\t\al^2}{1-\e^{-\t\al^2\N}}\)^{1/4}\Mpl
 =3.2\times 10^{16}r_\star^{1/4}\; \rm GeV\, ,
 \label{eq:scal_inf}
\end{equation}
which  for $r_\star=0.15$ gives us $V_{\rm in}^{1/4}=2\times 10^{16}\; \rm GeV$.
For this scale of inflation using Eq.~(\ref{Vratio}) and 
$\N=60$, $r_\star=0.15$ and $\t\al=0.06$, we find that
\begin{equation}
\bigg|\frac{\phi_{\rm end}}{\Mpl}\bigg|\approx \frac{24}{\al}\, ,
\end{equation}
which implies that the field value at the end of inflation is 
super-Planckian for $20\lesssim\al\lesssim24$ and sub-Planckian for $\al>24$. 

We can also estimate the field value at the beginning of inflation
considering 
the above calculated scale of inflation and the fact that for small field 
approximation the canonical field $\sig$ is related to the non-canonical 
field $\phi$ through the relation $\sig\approx (\al/\t\al)\phi$
\cite{Hossain:2014xha},
which makes the potential of the same form but with different slope
$\t\al$. 
We obtain the following estimate
\begin{equation}
 \bigg|\frac{\phi_{\rm in}}{\Mpl}\bigg|=\frac{19.2}{\al} \, ,
\end{equation}
from which we deduce that the non-canonical field $\phi$ is always 
sub-Planckian for $\al\gtrsim 20$ but the canonical field $\sig$ 
is super-Planckian for $\t\al\ll 1$. Since in this case the inflaton 
rolls from smaller to larger values, the field value of the 
canonical field $\sig$ at the end of inflation will also be 
super-Planckian.

%%%%%%%%%%%%%%%%%%%%%%%%%%%%%%%%%%%%%%%%%%%%%%%%%%%%%%%%%%%%%%%%%%%%%%%%%%
\subsection{Canonical field with brane correction}
%%%%%%%%%%%%%%%%%%%%%%%%%%%%%%%%%%%%%%%%%%%%%%%%%%%%%%%%%%%%%%%%%%%%%%%%%%

It is interesting to examine the Lyth bound in the case of steep braneworld
inflation. In this case, the Friedmann equation is modified due to
high-energy corrections to Einstein equations on the brane as
\cite{Shiromizu:1999wj,Maartens:1999hf,Apostolopoulos:2005ff} 
\begin{equation}
H^2=\frac{\rho}{3 \Mpl^2}\left(1+\frac{\rho}{2\lambda_{\rm B}}\right),
\end{equation}
and the slow-roll parameters
read \cite{Maartens:1999hf}
\begin{eqnarray}
&&\epsilon=\epsilon_0\frac{1+V/\lambda_{\rm B}}{\left(1+V/2\lambda_{\rm B}\right)^2}
\label{eq:epsilon}\\
&&\eta=\eta_0\left(1+V/2\lambda_{\rm B}\right)^{-1},
\end{eqnarray}
where $\lambda_{\rm B}$ is the brane tension and  $\epsilon_0$ and
$\eta_0$ are standard slow-roll parameters.

In the high-energy limit, that is when $V\gg \lam_{\rm B}$, we have
$\epsilon,\eta\ll 1$ despite the fact that  $\epsilon_0,\eta_0$ are large.
The
number of e-folds in this case is related to $\delta \phi$ and the
slow-roll parameter,  given by the following relation:
\begin{equation}
\N=\frac{1}{\Mpl^2}\int_{\phi_{\rm end}}^{\phi_{\rm in}}
\frac{V}{V'}\left(1+\frac{V}{2\lambda_{\rm B}}\right){\rm d}\phi \, .
\label{eq:e-fold_RS}
\end{equation}
Therefore, in the high-energy limit ($V\gg \lam_{\rm B}$), using Eq.
(\ref{eq:epsilon})
we can re-write Eq.~(\ref{eq:e-fold_RS}) as
\begin{equation}
 \N\approx\frac{1}{\Mpl^2}\int_{\phi_{\rm end}}^{\phi_{\rm in}}
\frac{V}{V'}\frac{V}{2\lam_{\rm B}}{\rm d}\phi=\frac{1}{\Mpl}\int_{\phi_{\rm end}}^{\phi_{\rm in}}
\frac{1}{\sqrt{\ep}}\sqrt{\frac{V}{2\lam_{\rm B}}}{\rm d}\phi \, .
\label{eq:e-fold_RS1}
\end{equation}
From Eq.~(\ref{eq:e-fold_RS1}) we obtain the relation
\begin{eqnarray}
\label{lythb1}
 \N\lesssim \frac{\del\phi}{\Mpl}\sqrt{\frac{1}{\ep_{\rm min}}\frac{V_{\rm max}}{2\lam_{\rm B}}}=
 \frac{\del\phi}{\Mpl}\sqrt{\frac{1}{\ep_{\rm min}}\frac{V_{\rm in}}{2\lam_{\rm B}}} \, ,
 \label{eq:e-fold_RS2}
\end{eqnarray}
where $\del\phi=|\phi_{\rm in}-\phi_{\rm end}|$. Hence, an interesting
expression for the Lyth bound on the brane follows from Eq.~(\ref{lythb1}) as
\begin{equation}
\label{lythb2}
 \delta \phi\gtrsim
\Bigg(\N\Mpl\sqrt{\frac{r_\star}{8}}\Bigg)\sqrt{\frac{2\lambda_{\rm
B}}{3V_{\rm in}}} \, ,
\end{equation}
where we have considered the fact that for brane inflation $r_\star=24\ep_{\rm
in}$.
Since during steep braneworld inflation $V_{\rm in}\gg\lambda_{\rm B}$,
expression (\ref{lythb2}) tells us that $\delta \phi$ is suppressed by high
energy corrections and can still be sub-Planckian despite $r_\star$ being
large. This is related to the fact that the slow roll is not realized due to
the potential slope and the smallness of the curvature, but it is
facilitated by the high-energy corrections to the Einstein equations on the
brane. We shall
give concrete numerical estimates for a particular case in the
following discussion.

In the case of braneworld models of quintessential inflation the
inflaton potential is a steep exponential, and only at late
times there is a feature in the potential which allows for an exit from the
scaling regime \cite{Sahni:2001qp,Sami:2004xk}. For instance a potential of
the form
\begin{equation}
V=V_0\Bigg[\cosh\(\frac{\gamma \phi}{\Mpl}\)-1\Bigg]^p,~~{\rm where}~~p>0 \, ,
\end{equation}
behaves like steep exponential for large $\phi$ provided that
$\alpha\equiv \gamma p$ is large, while around the origin $V\sim
\phi^{2p}$ such that the average equation of state is
$<\omega_\phi>=(p-1)/(p+1)$ and can give rise to the desired negative
value at late times. This scenario can describe late-time
acceleration provided that $V_0\sim 3H^2_0\Mpl^2$.

At early times, brane damping allows the field to derive inflation. We can
specialize to steep exponential potential, in which case 
(\ref{eq:e-fold_RS}) gives a simple relation, namely
  $V_{\rm in}=(\N+1)V_{\rm end}$ and  $V_{\rm in}/2\lam_{\rm B}=(\N+1)\al^2$ 
  \cite{Sahni:2001qp}. Additionally, the
Nucleosynthesis constraint \cite{Ade:2013zuv} leads to
\cite{Sahni:2001qp}
\begin{equation}
 \frac{\rho_\phi}{\rho_{\rm B}+\rho_\phi}=\frac{3(1+w_{\rm B})}{\al^2}\lesssim 0.01 \, ,
 \label{eq:neuc_cons_al}
\end{equation}
where $\rho_{\rm B}$ and $w_{\rm B}$ are respectively the background density
and equation-of-state parameter. For radiation, $w_{\rm B}=1/3$ and
from Eq.
(\ref{eq:neuc_cons_al}) we acquire  $\al\gtrsim 20$.

Finally, taking $\al=20,\; \N=60$ and $r_\star=0.1$ and using Eq.
(\ref{lythb2}), we get the bound on the range of inflation on
the brane, namely
\begin{equation}
 \del\phi\gtrsim \sqrt{\frac{r_\star}{24\(\N+1\)}}\frac{\N}{\al}\Mpl\simeq 0.0248 \Mpl \, .
 \label{eq:lyth_bound_RS}
\end{equation}
We can also independently estimate  $\del\phi$ for the exponential
potential, using the relation between $V_{\rm end}$ and $V_{\rm in}$
and the brane tension, as
\begin{equation}
 \del\phi=\frac{\Mpl}{\al}\ln\(\N+1\) \, ,
\end{equation}
which for the quoted values of parameters gives $\del\phi=0.2\Mpl$,
consistently with Eq.~(\ref{eq:lyth_bound_RS}).

Proceeding further, we can use relation $r_\star=24\ep_{\rm in}$, in the high-energy
limit ($V\gg
\lam_{\rm B}$), in order to estimate the scale of inflation in terms of the
tensor-to-scalar ratio as
\begin{equation}
 V_{\rm in}^{1/4}=\(\frac{96\al^2\lam_{\rm B}}{r_\star}\)^{1/4} \, .
 \label{eq:scale_inf_rs}
\end{equation}
The value of $\lam_{\rm B}$ is fixed from the
COBE normalization and is given by \cite{Sahni:2001qp}
\begin{equation}
 \lam_{\rm B}=\frac{2.6\times 10^{-10}}{\al^6}\(\frac{8\pi\Mpl}{\N+1}\)^{4} \, .
\end{equation}
Thus, for $\al=20,\; \N=60$ and $r_\star=0.4$  from  
(\ref{eq:scale_inf_rs}) we obtain
\begin{equation}
 V_{\rm in}^{1/4}=7.9\times 10^{14}\rm GeV \, .
\end{equation}

Let us note that for $\N=60$ and considering scale of inflation $\sim
10^{15}$ GeV we acquire
\begin{eqnarray}
&&\bigg|\frac{\alpha \phi_{\rm in}}{\Mpl}\bigg|\simeq
\ln \left(\frac{10^{60}}{3H^2_0\Mpl^2}\right)\simeq 244 \\
&&\bigg|\frac{\alpha \phi_{\rm end}}{\Mpl}\bigg|\simeq
\ln\left(\frac{10^{60}}{\(\N+1\)3H^2_0\Mpl^2}\right)\simeq 240,
\end{eqnarray}
which clearly show that inflation takes place in the range of large
$\phi$ for viable values of $\alpha$. The inflationary potential
contains a free parameter $\alpha$ to be fixed by post-inflationary
requirements. The requirement of scaling behavior implies that
$\alpha > \sqrt{3}$, but the nucleosynthesis constraint \cite{Ade:2013zuv}
demands larger numerical values, namely $\alpha\gtrsim 20$, such that
$\delta \phi <\Mpl$. 
In this case the Lyth bound does not impose any
restriction on the post-inflationary dynamics. On the other hand, it
is the post-inflationary dynamics which makes the range of inflation
sub-Planckian and the Lyth bound is evaded, despite the field values
being super-Planckian.\footnote{However, the tensor-to-scalar ratio
$r$ is slightly higher in the case of steep braneworld inflation.
Replacing the exponential potential by an inverse power-law
$\phi^{-p}$ makes the situation worse. In this case the minimum of the
ratio is reached for large $p$, as the inverse power law potential
approaches the exponential form. }

%%%%%%%%%%%%%%%%%%%%%%%%%%%%%%%%%%%%%%%%%%%%%%%%%%%%%%%%%%%%%%%%%%%%%%%%%%%%%%%
\section{conclusion and discussion}
%%%%%%%%%%%%%%%%%%%%%%%%%%%%%%%%%%%%%%%%%%%%%%%%%%%%%%%%%%%%%%%%%%%%%%%%%%%%%%%

In this letter we have examined the structure of the Lyth bound in two
classes of models of quintessential inflation. The quintessential inflation 
with non-canonical kinetic term can give rise to large tensor-to-scalar ratio
of perturbations, consistent with BICEP2 measurements \cite{Ade:2014xna},
such that the scale of inflation is around $10^{16} \rm GeV$.
The coefficient
of the kinetic term ($k^2(\phi)$) is given by relation~(\ref{eq:k}), which
has a maximum value $(\al/\t\al)^2$ for small field values
\cite{Hossain:2014xha} and approaches  unity in large-field approximations.
The parameter $\t\al$ controls the inflationary dynamics such that $\t\al\ll
1$, while $\al$ is related to post inflationary evolution and is constrained
by the nucleosynthesis bound \cite{Ade:2013zuv}. 
It should be noted that the parameter $\al$ does not appear 
in any physical quantity related to inflation. However, the presence of a 
non-canonical term in the action~(\ref{eq:action1}) modifies the Lyth bound
to   $\del\phi\gtrsim \frac{\N\Mpl}{k_{\rm max}}\sqrt{r_\star/8}$ 
with $k_{\rm max}=\al/\t\al$.  Since $\t\al\ll1$ and $\al\geq20$, we find
that $k_{\rm max}$ can be a large number leading to suppression of 
$\del\phi$. For instance, for a viable choice of parameters, $\N=60$,
$r_\star=0.15$, $\al=20$ and $\t\al=0.06$, we find that  $\del\phi\simeq
0.0246\Mpl$, which implies that the Lyth bound is clearly evaded in this
case. It is interesting to note that we have not imposed any extra
restriction other than the viability of inflation and post-inflationary
evolution.

In case of steep braneworld inflation, for $\N=60,\; \al=20\; {\rm and}\;
r_\star=0.1$ , we obtained $\del\phi=0.25\Mpl$, which obeys the Lyth bound and has 
sub-Planckian value. We have also calculated the field value at the beginning
and at the end of inflation.  We found that the non-canonical field, from the
beginning of inflation to its end, remains sub-Planckian provided  $\al\geq
24$, which is consistent with the nucleosynthesis constraint   $\al\geq 20$.
It is important to note that though the parameter $\al$ does not play any
role during inflation, it appears in the Lyth bound and plays a crucial role.
Nucleosynthesis constraint makes $\al$ large which helps to evade the Lyth
bound by making $\del\phi$ sub-Planckian.

In the steep braneworld inflation the slow roll takes place along a steep
potential due to the brane damping. The ratio $r$ is enhanced by a factor of
$V/\lambda_{\rm B}$ such that the Lyth bound is suppressed by its inverse, in
high-energy approximation valid during inflation. Similar to the
non-canonical case, nucleosynthesis constraint makes the range of inflation
$\delta \phi$ sub-Planckian on the brane, despite $r$ being  large. The scale
of steep braneworld inflation is about $10^{15}$ GeV. We should however
emphasize that inflation is realized for super-Planckian values of the field
in this case. Secondly, the induced numerical value of $r$ is slightly higher
than the observed values in the BICEP2 experiment \cite{Ade:2014xna}.

It is indeed interesting that the Lyth bound imposes severe restrictions on
the single canonical scalar field models of inflation in the standard FRW
cosmology. However, as we demonstrated, the bound can be evaded in case of
models of quintessential inflation considered here. Unfortunately, the steep
braneworld inflation cannot meet the BICEP2 requirement.

 It is remarkable that the class of models corresponding to action~(\ref{eq:action1}) not only evades the Lyth bound but also meet the BICEP2
requirements.  We have therefore demonstrated that $\delta \phi$ is 
sub-Planckian despite the tensor-to-scalar ratio of perturbations $r$ being
large.
 
An important comment relating to the effective field theoretic description 
of inflation is in order. As we mentioned earlier, the canonical field
$\sigma$ associated with $\phi$ is super-Planckian. We might take the
orthodox view and abandon the transformation to venture into super-Planckian
region and focus on the non-canonical-field ($\phi$) description. But in that
case, unlike the canonical Lagrangian where one computes quantum correction
to potential, the corrections should be calculated to the total Lagrangian
\cite{Baumann:2011ws,Cheung:2007st,Baumann:2014nda}. As noticed in the case
of polynomial type of Lagrangian, these correction might get large for large
values of the kinetic function $k(\phi)$. This could possibly be the
manifestation of the problem shifted from canonical Lagrangian to its
non-canonical description, and it might be worthwhile to check it for the
Lagrangian considered here. The full investigation of the quantum behavior
lies beyond the scope of the present letter, and is left for a future
project.

\begin{acknowledgments}
MWH acknowledges CSIR, Govt. of India for financial support through
SRF scheme. The research of ENS is implemented within the framework of the
Operational Program ``Education and
Lifelong Learning'' (Actions Beneficiary: General Secretariat for Research
and Technology), and is co-financed by the European Social Fund (ESF) and the
Greek State. MS thanks the Eurasian  International
Center for Theoretical Physics, Astana for hospitality where the work was initiated.
\end{acknowledgments}

\end{document}